\documentstyle[aps]{revtex}   

\begin{document}   
   
\twocolumn[\hsize\textwidth\columnwidth\hsize\csname@twocolumnfalse%
\endcsname   
\draft   
   
\title{   Radiating dipoles in photonic crystals}   
   
\author{  Kurt Busch$^{1,2}$, Nipun Vats$^1$,    
          Sajeev John$^1$ and Barry C.\ Sanders$^3$}   
   
\address{ $^1$ Department of Physics, University of Toronto,   
	       60 St.\ George Street, Toronto, Ontario, Canada M5S 1A7\\   
          $^2$ Institut f\"ur Theorie der Kondensierten Materie,   
               Universit\"at Karlsruhe, P.O. Box 6980, 76128 Karlsruhe, Germany\\
       	  $^3$ Department of Physics, Macquarie University,   
	       Sydney, New South Wales, Australia 2109 }   
   
\date{Revised: \today}   
   
\maketitle   
   
\begin{abstract}   
   
The radiation dynamics of a dipole antenna embedded in a Photonic Crystal   
are modeled by an initially excited harmonic oscillator coupled to a    
non--Markovian bath of harmonic oscillators representing the colored 
electromagnetic vacuum within the crystal.  
Realistic coupling constants    
based on the natural modes of the Photonic Crystal, i.e., 
Bloch waves and    
their associated dispersion relation, are derived. For simple model systems,    
well-known results such as decay times and emission spectra are reproduced.   
This approach enables direct incorporation of realistic band structure    
computations into studies of radiative emission from atoms and molecules 
within photonic crystals. We therefore provide a predictive and 
interpretative tool for experiments in both the microwave and optical 
regimes.      
   
\end{abstract}   
   
   
   
]   
   
\narrowtext   
  
\section{Introduction} 
   
Photonic crystals (PCs) have been the subject of intensive research over   
the past decade\cite{Sou96}.  
These are fabricated periodic dielectric arrays that employ a combination  
of (i)~Mie scattering from individual elements of the array, and (ii)~Bragg  
scattering from the periodic lattice, to induce a band structure for photon  
propagation. This band structure is, in many ways, analogous to electronic  
band structure in a semiconductor. Through a judicious selection of materials  
and of the periodicity of the lattice, the photonic dispersion relation and  
the associated electromagnetic (EM) mode structure of a PC can be adapted to  
a variety of device applications. The most dramatic modification of the  
photon dispersion occurs when the linear propagation of a photon in a PC is  
prohibited in all directions for a range of frequencies, giving rise to a  
complete photonic band gap (PBG).    
    
The radiative dynamics of an optically active material placed within  
or near a PC can be dramatically modified from free space.  This is a  
result of the ``colored'' electromagnetic reservoir provided by the  
solutions to the electromagnetic field equations within a PC. In the optical  
domain, theoretical studies of atomic transitions coupled to the EM modes of  
a PC with an optical PBG predict a number of novel quantum optical  
phenomena. These phenomena include: the suppression or enhancement of  
spontaneous emission and the associated fractional localization of light  
near radiating atoms \cite{Joh91,Qua94}; rapid all--optical switching 
\cite{Qua97}; and anomalous superradiant emission, as well as low--threshold  
lasing near the edge of a PBG\cite{Qua95,Vat98}. Microfabrication  
of PCs with complete PBGs at optical wavelengths has proven to be a difficult  
task, both because the lattice periodicity should be comparable to the  
wavelength of the light under consideration, and because a high dielectric  
contrast between the elements of the lattice is required.   
Fortunately, recent advances in microlithography \cite{Lin99}and in 
semiconductor   
infiltration in colloidal crystals \cite{Wij98} have produced materials with   
significant pseudo--gaps in their photonic band structure \cite{Thi99}.   
The development of materials with complete PBGs in the optical regime  
appears imminent.   
  
High--quality PBG materials at microwave frequencies have been available for  
some time \cite{Ozb95}. Sizeable band--gaps with center frequencies ranging  
from a few GHz up to 2 THz have been reported; these crystals have thus proven  
the soundness of the concept of the PBG. Microwave PBG materials may be  
relatively easily manufactured using micro-machining techniques, and are  
currently of interest for applications such as the shielding of human tissue  
from microwave radiation, and for improving the radiation characteristics of  
microwave antenn\ae . Although PBG materials at microwave frequencies 
have been extensively  
studied, the behavior of radiating dipolar antenn\ae~  
embedded in microwave PCs has not received the same degree of attention. 
This is despite the fact that such antenn\ae~ would share many properties  
in common with atomic emission in a PC. In the microwave domain, a dipole  
antenna could take the form of an electrically excited metallic pin with  
a high~$Q$ (quality) factor.  
 
The radiative dynamics of the above system can be  
described by a charged, one--dimensional simple harmonic oscillator (SHO).     
Such an electric dipole oscillator can also provide an excellent  
description of the radiation of single or multiple two--level atoms in the  
optical domain. This description is valid provided that the total  
excitation energy of the atoms is well below an energy where saturation  
(nonlinear) effects become important. Moreover, the radiation reservoir can  
itself be modeled as a bath of many independent SHOs: Radiative damping  
arises from a linear coupling between the system SHO and the large number of  
reservoir oscillator modes.  
The similarities between the microwave and optical systems, coupled  
with the mature state of microwave technology, suggests that many of the  
predicted effects for atomic dipoles in the optical domain could be realized  
and studied first in the microwave domain.  
 
Analytical techniques exist for treating certain forms of coupling  
between the dipole and reservoir for certain modal distributions of  
the reservoir.  However, PCs present coupling distributions and spectral  
properties which defy analytical methods.   
This is due to the presence of a restricted and rapidly--varying reservoir  
mode distribution, which renders invalid the usual Born--Markov type 
approximation schemes for the system--reservoir interaction.  
To obtain accurate results, we solve the system numerically for a large,  
but finite, number of oscillators in the reservoir by discretizing  
the modes of the reservoir following the approach of Ullersma \cite{Ull66}.  
In dealing with our system, there are crucial issues concerning obtaining  
the correct coupling strength between the oscillator and the reservoir  
modes, as well as in employing the proper renormalization and mode sampling  
in numerical simulations. When these criteria are satisfied, the SHO method  
comprises a powerful approach to treating radiative dynamics.  
  
Here, we develop a rigorous quantitative treatment of the radiative dynamics  
of an electric dipole oscillator coupled to the electromagnetic reservoir  
within a model PC. In the process, we provide a sound theoretical 
basis for this and other approaches \cite{Nik99} to non-Markovian radiative 
dynamics which involve the discretization of a model electromagnetic 
reservoir.  Additionally, we show how our method can be applied to 
realistic PC's with 
complicated dispersion relations and EM mode structures. 
The paper is organized as follows.  
In Section~\ref{sec:classical}, we develop a classical field theory for  
electromagnetic field modes in PCs, and we derive the coupling constants for  
the coupling between a radiating dipole and these Bloch modes. This 
leads to the 
Hamiltonian of the coupled system and the associated equations of motion. 
Renormalization issues arising from the non--relativistic nature of our  
theory are discussed in Section III, whereas Section IV describes the  
discretization of the reservoir and the numerical solution of the equations  
of motion. In Section V, these techniques are applied to a highly  
computationally challenging model, that of a three--dimensional, isotropic  
dispersion relation with a complete PBG.  
The demonstration of fractional localization and related phenomena  
validates the SHO approach to modeling radiative dynamics in PCs.  
In Section VI we summarize the results and emphasize the possibilities  
for testing these predictions experimentally in the microwave domain.   
The two appendices are concerned with the details of the field theory  
for the PC and with the details of the model of the one--sided, isotropic  
PBG, respectively.  
   
\section{Classical field theory}  
\label{sec:classical}  
  
In this section, we derive the equations governing the dynamics of a  
radiating dipole oscillator located inside a PC.  
Typically the equation of motion for a damped oscillator, with time-dependent  
coordinate~$q(t)$, is written as the second--order differential equation    
\begin{equation}    
\label{secorder}    
\ddot{q}(t) + \gamma \dot{q}(t) + \omega_0^2 q(t) = F(t).    
\end{equation}    
Here, we have introduced a damping constant $\gamma$, the natural   
frequency $\omega_0$ and the driving field $F(t)$ for the amplitude $q$ of   
the linear oscillator. For instance, for a freely oscillating $RLC$ circuit   
with ohmic resistance $R$, capacitance $C$ and inductance $L$, we have   
$\gamma = R/L$, $\omega_0^2 = 1/LC$, $F(t) =0 $, and $q(t)$  is the electric  
charge.    
Eq.\ (\ref{secorder}) is, however, not the most general way of incorporating   
damping into the equations of motion for a harmonic oscillator.  
This description can break down if, for example, there is a suppression of  
modes in the reservoir to which the dipole oscillator can couple.  
Such a suppression of modes is a feature of the EM reservoir present in a PC.  
A more general description of damping forces acting on the harmonic  
oscillator therefore requires a precise knowledge of the mode structure of  
its environment, and the corresponding coupling of the system oscillator to  
these modes. In the case of a radiating dipole located in a PC, it is then  
appropriate to model its emission dynamics with a SHO coupled to a 
reservoir of SHOs.  
The essential difference between the vacuum and a PC is then contained in  
the spectral distribution, or density of states (DOS), of the reservoir  
oscillators, and in the coupling constants between the reservoir modes and the  
system oscillator.     
   
The characterization of the reservoir is carried out in detail in Appendix   
A; here we only summarize the salient results.  
Given a radiating dipole with a natural frequency $\omega_0$, we obtain the  
classical Hamiltonian 
\begin{equation}    
\label{H}    
H = H_{\rm dip} + H_{\rm res} + H_{\rm ct} + H_{\rm int}~~ .  
\end{equation}    
The first term on the right--hand side of the Hamiltonian  
is the energy of the dipole oscillator itself,  
\begin{equation}    
\label{Hdip}    
H_{\rm dip} = \xi \, \omega_0 \, \vert \alpha \vert^2 .  
\end{equation}    
The natural frequency of the isolated oscillator is~$\omega_0$,  
and~$\xi$ is a constant with the dimension of energy $\times$ time. 
This permits us to write the energy of a SHO in units of   
its natural frequency $\omega$, {\em i.e.}, $E(\omega ) = \xi\omega$. 
The system oscillator's complex amplitude is given by the dimensionless,   
time--dependent quantity $\alpha$, defined with respect to the  
coordinate~$q(t)$ of Eq.\ (\ref{secorder}) as  
\begin{equation}   
\alpha(t) \equiv   \sqrt{\frac{L \omega_0}{2 \xi}} q(t) +    
            \imath \sqrt{\frac{1}{2 \xi L \omega_0}} (L \dot{q}(t)) ~~.  
\end{equation}   
 
The next term in the Hamiltonian~(\ref{H}) corresponds to the free  
evolution of the radiation reservoir, which is modeled as a bath of 
independent SHOs,  
\begin{equation}  
H_{\rm res}  = \sum_\mu \xi \, \omega_\mu \, \vert \beta_\mu \vert^2 .   
\end{equation}   
The natural electromagnetic modes of the PC are Bloch modes (see Appendix   
A), labeled with the index $\mu \equiv (n \vec{k})$, where $n$ stands for   
the band index and $\vec{k}$ is a reciprocal lattice vector that lies in   
the first Brillouin zone (BZ). Their dispersion relation, $\omega_\mu$, is  
different from the vacuum case, and may have complete gaps and/or the  
corresponding density of states may exhibit appreciable pseudogap structure,  
the manifestation of multiple (Bragg) scattering effects in periodic media.   
  
As we are working within the framework of a non-relativistic field theory,  
we have introduced a mass renormalization counter term  
$H_{\rm ct} = - \xi\Delta \vert \alpha \vert^2$   
that cancels unphysical UV-divergent terms \cite{All87,Bet47}.  
The quantity~$\Delta$ is specified in Section III.  
  
The interaction between the oscillator and the reservoir is given by a linear  
coupling term. As the oscillator frequency is quite large, and the effective  
linewidth of the oscillation is relatively small, it is possible to simplify  
the interaction by applying the rotating--wave approximation.  
In this approximation, couplings in the Hamiltonian of the  
form $\alpha^* \beta_\mu^*$ and its complex conjugate are neglected,  
as these terms oscillate very rapidly compared to the terms  
of the type $\alpha^* \beta_\mu$ and its conjugate.  
Hence, the interaction Hamiltonian can be expressed as  
\begin{equation}  
\label{H:interaction}  
H_{\rm int} =	 - \imath \xi\sum_\mu    
          \left( \, \alpha^* \, g_\mu^* \, \beta_\mu -   
                    \alpha \, g_\mu   \, \beta_\mu^*    
          \right)~~.  
\end{equation}   
   
In the case of a point dipole, i.e., when its spatial   
extent $a$ is much smaller than the wavelength corresponding to its   
natural frequency, $\lambda_0 = 2 \pi   
\omega_0/c$, the coupling constants $g_\mu$ can be derived from  
(i)~the magnitude of the dipole moment, $d(t) = a q(t)$, located at 
$\vec{r}_0$, and  
(ii)~the dipole orientation, $\hat{d}$, relative to that of   
the Bloch modes, $\vec{E}_\mu (\vec{r}_0)$:   
\begin{equation}   
\label{Coupling}   
g_\mu \equiv g_\mu(\vec{r}_0) =    
ac \sqrt{\frac{ \pi}{L \omega_0 \omega_\mu}}   
        \, \left(\hat{d} \cdot \vec{E}_\mu^*(\vec{r}_0) \right).   
\end{equation}   
This dependence of the coupling constant on the dipole's precise location   
within the PC is the second essential difference from the   
free--space case. As shown in Refs.\ \cite{Bus98,Spr96}, this position   
dependence may be quite strong, thus making its incorporation a {\em   
sine qua non} for any quantitative theory of of radiating antenn\ae~ or   
fluorescence phenomena in realistic PCs.   
   
The emission dynamics can be evaluated from the Poisson brackets of the    
oscillator amplitudes and their initial values, $\alpha(0) = 1$ and   
$\beta_\mu(0) = 0$ $(\forall \mu)$. Our choice of    
$\alpha(0)$ and $\beta_\mu(0)$ corresponds to the initial condition of an    
excited dipole antenna and a completely de--excited bath. The only  
non--zero Poisson brackets are   
\begin{equation}   
\label{Poisson}   
\left\{ \alpha, \alpha^* \right\} = \left\{ \beta_\mu, \beta_\mu^* \right\} = 
\frac{\imath}{\xi}.   
\end{equation}    
  
Eqs.\ (\ref{H}), (\ref{Coupling}) and (\ref{Poisson}),  
together with the initial values for the oscillator amplitudes,  
completely determine the emission dynamics of a radiating dipole embedded  
in a PC. In the following sections, we solve the corresponding   
equations of motion. This task is complicated by the nature of the  
reservoir's  excitation spectrum: as discussed, the non-smooth density of  
states prohibits the use of a Markovian approximation and its appealing  
simplifying features \cite{Joh91,Qua94,Vat98}. Instead, we have to revert  
to a solution of the full non--Markovian problem. This is accomplished by  
firstly rearranging the reservoir modes in a manner more suitable to both  
analytical as well as numerical solutions, and subsequently solving the  
equations of motion.  In what follows, we bridge the gap between  
previous studies of simplified model dispersion relations   
\cite{Joh91,Qua94,Vat98} and band structure computations   
\cite{Bus98,Ho90}.  
 
Although we will formally develop our theory for an LC circuit in a microwave  
PC, we emphasize that the formalism applies equally well to a semiclassical  
Lorentz oscillator model of an excited two-level atom, i.e., an electron with  
charge~$e$ and mass~$m$ which is bound to a stationary nucleus, for which  
the energy of excitation is well below that required for saturation 
effects to become relevant.  
The oscillator coordinate~$q(t)$ may then be identified with the deviation of  
the electron's position from its equilibrium value, $\gamma$ is the inverse  
life time of the excited state, and $\omega_0$ denotes the frequency for  
transitions between excited and ground state of the two-level atom.  
This corresponds to making the substitutions:  
\begin{equation}   
L \to m, ~~ (L \dot{q}) \to p, ~~ \xi\to \hbar,   
\end{equation}   
where~$h = 2 \pi \hbar$ is Planck's constant.

\section{Projected Local Density of States, Mass renormalization and   
Lamb shift}   
   
From the Hamiltonian (\ref{H}) we derive the equations of motion for the  
amplitudes    
\begin{eqnarray}   
\label{EqofM1}   
\dot{\alpha}(t) & = & - \imath \left(\omega_0-\Delta\right) \, \alpha(t)   
                      - \imath \, \xi \sum_\mu g_\mu^* \, \beta_\mu (t)\\   
\label{EqofM2}   
\dot{\beta_\mu}(t) & = & -\imath \, \omega_\mu \, \beta_\mu(t)     
                        +g_\mu \, \alpha(t),   
\end{eqnarray}   
for which we seek a solution with initial conditions $\alpha(0) = 1$ and   
$\beta_\mu(0) = 0$ ($\forall \mu)$. Our formalism however requires that   
we first determine the mass renormalization counter term $\Delta$. This is   
most conveniently done in a rotating frame with slowly varying amplitudes   
$a(t)$ and $b(t)$, defined as $\alpha(t) = a(t) e^{-\imath\omega_0t}$ and   
$\beta(t) = b(t) e^{-\imath\omega_\mu t}$ respectively:   
\begin{eqnarray}   
\label{EqofMRot1}   
\dot{a}(t) & = & - \imath \, \xi \sum_\mu g_\mu^* \,    
                         e^{\imath(\omega_0-\omega_\mu)t} \, b_\mu (t)
                         + \imath \Delta a(t)\\   
\label{EqofMRot2}   
\dot{b}(t) & = & g_\mu \, e^{-\imath(\omega_o-\omega_\mu)t} \,   
                         a(t).   
\end{eqnarray}   
Conversely, Eqs.\ (\ref{EqofMRot1}) and~(\ref{EqofMRot2}) comprise  
a stiff set of differential equations making their solution a 
difficult task. Numerical solution of the problem 
is more easily performed in the non-rotating frame, to which we return 
in Sect. IV.   
   
Eq.\ (\ref{EqofMRot2}) may be formally integrated,    
\begin{equation}   
b_\mu(t) = g_\mu \int_0^t dt^\prime \,   
           e^{-\imath(\omega_0-\omega_\mu)t^\prime} \,   
           a(t^\prime),   
\end{equation}   
and inserted into Eq.\ (\ref{EqofMRot1}) to yield   
\begin{equation}   
\label{EqofMGreen}   
\dot{a}(t) = - \int_0^\infty dt^\prime \, G(t-t^\prime) \, a(t^\prime)    
             + \imath \Delta a(t),   
\end{equation}   
where the Green function $G(\tau)$ contains all the information   
about the reservoir and is the subject of our studies for the remainder   
of this section.  It is defined as   
\begin{equation}   
\label{Green}   
G(\tau) \equiv \Theta(\tau) \sum_\mu \, \vert g_\mu \vert^2 \,    
                e^{\imath(\omega_0-\omega_\mu)\tau} ~~.   
\end{equation}   
Here, $\Theta(\tau)$ denotes the Heaviside step function, which  
ensures the causality of $G(\tau)$.   
We now proceed to evaluate $G(\tau)$ for the form of    
the coupling constants $g_\mu$ given in Eq.\ (\ref{Coupling}).    
To this end, we introduce the projected local DOS (PLDOS)   
$N(\vec{r}_0,\hat{d}, \omega)$ through   
\begin{equation}   
\label{LDOS}   
N(\vec{r}_0,\hat{d},\omega) = \sum_n \int_{\rm BZ} \frac{d^3k}{(2\pi)^3} \,   
                    \delta(\omega-\omega_{n\vec{k}}) \,   
                    \vert \hat{d} \cdot \vec{E}_{n\vec{k}}(\vec{r}_0)    
                    \vert^2 ,   
\end{equation}   
where we have replaced the symbolic sum over $\mu$ by its proper   
representation as a sum over bands plus a wave vector integral over   
the BZ. With these changes, we may rewrite $G(\tau)$ compactly as   
\begin{equation}   
\label{GreenLDOS}   
G(\tau) = \beta \, \Theta(\tau) \int_0^\infty d\omega \,    
                             \frac{N(\vec{r}_0,\hat{d},\omega)}{\omega} \,    
                             e^{\imath (\omega_0 - \omega) \tau} .   
\end{equation}   
Here, we have abbreviated $\beta = (\pi a^2 c^2)/(L \omega_0)$.    
Eq.\ (\ref{GreenLDOS}) makes more explicit what we have argued before:   
The spontaneous emission dynamics of active media in Photonic Crystals   
are completely determined by the PLDOS, $N(\vec{r}_0,\hat{d},\omega)$. As    
the PLDOS may be drastically different from location to location   
within the Wigner--Seitz cell of the PC \cite{Bus98,Spr96},    
it is imperative to have detailed knowledge about where in the PC the   
dipole is situated in order to understand and predict the outcome of  
corresponding experiments.   
   
One additional point deserves special attention: the total DOS,  
$N(\omega)$, is related to the local DOS via   
\begin{eqnarray}   
N(\omega) & = & \frac{1}{V} \int_V d^3r \, \int d \Omega_{\hat{d}} \,  
                \epsilon_p (\vec{r}) \,    
                N(\vec{r},\hat{d},\omega) \nonumber \\   
          & \ne & \frac{1}{V} \int_V d^3r \int d \Omega_{\hat{d}}   
            \, N(\vec{r},\hat{d},\omega) \nonumber ,   
\end{eqnarray}   
where $V$ is the volume of the Wigner--Seitz cell, and  
$\int d \Omega_{\hat{d}}$ is the average over all possible orientations of  
the dipole. Strictly speaking, it is not possible to base conclusions about  
the radiation dynamics on the total DOS. This is a direct consequence of the  
fact that the natural modes of PCs are Bloch waves rather than plane waves  
as in free space.  Depending on the band index, these Bloch modes prefer to   
``reside'' predominantly in either low or high dielectric index regions  
(so-called air and dielectric bands respectively). Only in the case of very  
low index contrast (``nearly free photons'') may the total DOS be viewed as  
a reliable guide to interpreting radiative dynamics within a PC. The total  
DOS is, nevertheless, an adequate rule-of-thumb estimator.   
  
From Eq.\ (\ref{LDOS}) we can now obtain the Fourier transform of the Green  
function, $G(\Omega-\omega_0)$, centered around the atom's bare transition  
frequency $\omega_0$:   
\begin{eqnarray}   
G(\Omega-\omega_0) & = & \int_0^\infty dt \,    
                        e^{\imath(\Omega - \omega_0) t} \,   
                        G(t) \nonumber \\   
                  & = & \pi \beta \, \frac{N(\vec{r}_0,\hat{d},\Omega)}{\Omega}                          \, \Theta(\Omega) \nonumber \\   
                  &   &  + \, \imath \beta \int_0^\infty d\omega \,   
                          \frac{N(\vec{r}_0,\hat{d},\omega)}{\omega} \,   
                          \wp \left( \frac{1}{\Omega-\omega} \right) ,   
                          \nonumber   
\end{eqnarray}   
where $\wp$ stands for the principal value.   
   
For large $\omega$, we have $N(\vec{r}_0,\hat{d},\omega)    
\propto \omega^2$.  
The imaginary part of $G(\Omega-\omega_0)$ apparently contains a linear   
divergence in the UV. This divergence is to be expected for a non-relativistic 
theory, analogous to the problem of spontaneous emission in vacuum    
\cite{All87}, and is removed from the theory by using the mass counter  
renormalization term $\Delta$, as first pointed out by    
Bethe \cite{Bet47}.  
Consequently, we decompose the imaginary part   
of $G(\Omega-\omega_0)$ into   
\begin{equation}   
   \Im \left( G(\Omega-\omega_0) \right) \simeq    
       - ( \Delta + \delta_{\rm vac} + \delta_a ),   
\end{equation}   
where we have used the notation:   
\begin{eqnarray}   
   \Delta & = & \beta \int_0^\infty d\omega \,    
                \frac{N(\vec{r}_0,\hat{d},\omega)}{\omega^2}  \nonumber \\   
   \delta_{\rm vac} & = & - \frac{\beta \, \omega_0}{\pi^2 c^3}   
                      \int_0^{\Omega_c} d\omega \,    
                      \wp \left( \frac{1}{\omega_0-\omega} \right)    
                      \nonumber \\   
   \delta_{\rm a} & = & - \frac{\beta \, \omega_0}{\pi^2 c^3}   
                      \int_0^{\Omega_c} d\omega \,    
                      \wp \left( \frac{1}{\omega_0-\omega} \right) \times
                      \nonumber \\   
                  &   & ~~~~~~~~~~~~~~~~ \times
                      \frac{N(\vec{r}_0,\hat{d}, \omega)-N^{\rm (vac)}(\omega)}
                           {\omega^2}   
                      \nonumber .   
\end{eqnarray}   
Here, we have performed a Wigner--Weisskopf-type approximation on the vacuum  
and anomalous Lamb shifts \cite{Joh91}, $\delta_{\rm vac}$ and $\delta_{\rm a}$,
respectively. This approximation is  
justified by the fact that, despite its highly non--Markovian nature, a  
radiating dipole in a PC is still a weak coupling problem,      
as can be seen, for instance, by estimating the coupling constant    
\begin{equation}  
g \simeq d_0 \omega_0 \sqrt{\frac{2 \pi}{ V \xi \omega_{n\vec{k}}}}  
\end{equation}  
in the Lorentz oscillator model.  
Here, $V \approx \bar{a}^3$ is the volume of    
the Wigner--Seitz cell     
of the PC ($\bar{a}$ is the corresponding lattice constant)    
and $d_0 = e a_0$ is the oscillator's dipole moment for the elementary  
charge $e$ and Bohr atomic radius, $a_0$. At optical frequencies  
($\omega \approx 10^{15}$ s$^{-1}$), a silicon inverted opal has a PBG at  
the frequency $\bar{a} \omega / 2 \pi c \approx 0.8$, so that we obtain   
$10^{-7} \le g/\omega_0 \le 10^{-6} \ll 1$, thus justifying our   
Wigner--Weisskopf approximation. As a consequence, we  
must treat the real part of $G(\Omega-\omega_0)$ exactly, but are still  
allowed to tackle the imaginary part of $G(\Omega-\omega_0)$ using standard  
perturbation methods of QED.   
In addition, we have introduced the vacuum or free--space DOS   
$N^{\rm (vac)}(\omega) = \omega^2/(\pi^2 c^3)$, and a cutoff frequency   
$\Omega_c \gg \omega_0$, which is chosen large enough that   
the results of the following analysis remain independent of the precise   
value of $\Omega_c$.  
In a Lorentz oscillator model, for instance,   
$\Omega_c$ can be identified with the Compton frequency    
$\Omega_c \simeq m c^2 / \hbar $, as $\omega > \omega_c$   
probes the relativistic aspects of the oscillating charge, which are beyond  
the scope of the model.   
   
With the foregoing analysis, we have determined the mass renormalization   
counter term $\Delta$. In addition, we have derived an explicit expression  
for the anomalous Lamb shift $\delta_a$ \cite{Joh91} which originates in  
the ``reshuffling'' of the reservoir's spectral weight by the PC.  
   
\section{Discretization of the reservoir}   
   
To solve the equation of motion for the amplitude of the system oscillator,   
let us rewrite Eq.\ (\ref{EqofMGreen}) in a more explicit form:   
\begin{eqnarray}   
\label{EqofMExp}   
\dot{a}(t) & = & - \int_0^\infty \! d\omega \, N(\vec{r}_0,\hat{d},\omega) \,    
                   g^2(\omega) \int_0^t \! dt^\prime  \,   
                   e^{\imath(\omega_0-\omega)(t-t^\prime)} a(t^\prime)   
                   \nonumber \\    
           &   &   + \imath \Delta \, a(t) ,   
\end{eqnarray}   
where $g^2(\omega)=\beta/\omega$, and the mass renormalization counter term    
$\Delta$ is given by   
\begin{equation}    
\Delta = \beta \int_0^\infty d\omega \, \frac{N(\vec{r}_0,\hat{d},\omega)}   
{\omega^2} .   
\end{equation}                
We remind the reader that $a(0)=1$.    
   
We are now in a position to comment on the origin of the linear damping term   
$\gamma \dot{q}(t)$ that appears in Eq.\ (\ref{secorder}): If we    
consider the long time limit, {\em i.\ e.}, $t \gg 1/\omega_0$, and 
assume that   
the PLDOS $N(\vec{r}_0,\hat{d},\omega)$ is a smooth function for frequencies    
around $\omega_0$, we can approximate the frequency integral in    
Eq.\ (\ref{EqofMExp}) by    
$\left[ 2 \pi \beta N(\vec{r}_0,\hat{d},\omega_0)/\omega_0\right] \delta(t -   
t^\prime)$, which leads to   
\begin{equation}   
\label{Markov}   
\dot{a}(t) =  - \gamma a(t) ,   
\end{equation}   
where the decay constant is defined as    
\begin{equation}  
\gamma = 2 \pi \beta N(\vec{r}_0,\hat{d},\omega_0) /\omega_0 .  
\end{equation}  
This approximation is is valid only for long times relative to  
$1/\omega_0$, and for a sufficiently smooth density of states. However,  
in the case of a PC, the PLDOS may have sharp discontinuities and   
gaps, thus requiring that the full equations of motion instead.   
  
To solve the integro-differential equation (\ref{EqofMExp}) in a PC, we   
appeal to the literal meaning of the PLDOS as a density of states:  
$N(\vec{r}_0,\hat{d}, \omega)$ may be interpreted as an unnormalized  
probability density of finding a reservoir oscillator with frequency   
$\omega$ at position $\vec{r}_0 $ and orientation~$\hat{d}$.  
Consequently, we transform Eq.\ (\ref{EqofMExp}) back to a system of  
coupled differential equations by employing a Monte Carlo integration scheme  
for an arbitrary function $f(\omega)$ according to    
\begin{eqnarray}   
\int_0^\infty d\omega \, N(\vec{r}_0, \hat{d}, \omega) \, f(\omega) & \simeq &   
\int_0^{\Omega_c} d\omega \, N(\vec{r}_0, \hat{d}, \omega) \, f(\omega)    
\nonumber \\    
& \approx & \frac{N_0}{M} \sum_{i=1}^M f(\omega_i)  ,   
\end{eqnarray}   
where the normalization constant    
\begin{equation}  
N_0 = \int_0^{\Omega_c} d\omega \, N(\vec{r}_0, \hat{d}, \omega)  
\end{equation}  
depends on the cutoff frequency, $\Omega_c$.  
There are $M\gg1$ bath oscillators, contained within a set of frequencies    
$\{ \omega_i, 1 \le i \le M \}$, the frequencies of which     
are obtained by randomly sampling the interval $ [ 0,\Omega_c ] $   
according to the probability density    
$p(\vec{r}_0,\hat{d}, \omega) = N(\vec{r}_0,\hat{d},\omega)/N_0$. Note  
that the $\omega_i$ may be degenerate, as prescribed by  
$p(\vec{r}_0,\hat{d}, \omega) $.  
   
Applying this Monte Carlo scheme to Eq.\ (\ref{EqofMExp}) and transforming   
back to a non-rotating frame in order to avoid having to solve a numerically  
stiff problem, we obtain   
\begin{eqnarray}   
\label{EqofMFin1}   
\dot{\alpha}(t) & = & - \imath \left(\omega_0-\Delta\right) \, \alpha(t)   
                      - \imath \, \xi \sum_{i=1}^N g_i \, \beta_i (t)\\   
\label{EqofMFin2}   
\dot{\beta_i}(t) & = & -\imath \, \omega_i \, \beta_i(t)     
                        +g_i \, \alpha(t) ,   
\end{eqnarray}   
where $g_i = g(\omega_i),~ 1 \le i \le M$, and the mass renormalization    
counter term is evaluated up to the cutoff frequency $\Omega_c$, {\em i.e.},   
$\Delta = \int_0^{\Omega_c} d\omega \, N(\vec{r}_0,\hat{d}, \omega)/\omega^2$. 
   
When comparing Eqs.\ (\ref{EqofMFin1}) and (\ref{EqofMFin2}) to our initial    
equations of motion, Eqs.\ (\ref{EqofM1}) and (\ref{EqofM2}), we observe   
that the considerations in the previous section have allowed us to    
rearrange the three-dimensional wave vector sum over the modes    
$\mu \equiv (n\vec{k})$ into a simple one-dimensional sum over a set of    
frequencies $\{ \omega_i\}$ with a probability distribution    
$p(\vec{r}_0,\hat{d}, \omega)$ that is easily determined through   
standard photonic band structure computation \cite{Bus98}. In the   
following section, we give the solutions    
of (\ref{EqofMFin1}) and (\ref{EqofMFin2}) for a model system    
which has previously been treated by other methods.  
In particular, we will demonstrate that known results for the radiative  
dynamics can be recaptured and do not depend on the the value of the cutoff  
frequency $\Omega_c$ and the number $M$ of reservoir oscillators once these  
quantities are large enough such that all the relevant features of  
$N(\vec{r}_0,\hat{d}, \omega)$, are adequately represented.   
   
\section{Numerical Results for a model system}   
  
In order to establish the validity of our approach, we now solve  
Eqs.\ (\ref{EqofMFin1}) and (\ref{EqofMFin2}) for a generic  
model of a PBG, the three-dimensional isotropic, one-sided PBG  
\cite{Qua94}.   
In Appendix B, we outline the construction of the model's dispersion relation   
and how to obtain the corresponding model DOS, $N_m(\omega)$.  We note  
that we do not appeal to an effective mass approximation in the  
dispersion relation \cite{Vat98}, as is done in most treatments of  
band--edge dynamics.  This allows us to recover the correct form of  
the large frequency behavior of the photon density of states.  
  
In Fig.\ 1,   
we show the behavior of $N_m(\omega)$ as a function of frequency  
for values of the relevant parameters,  
the gap size parameter $\eta=0.8$ and the normalized center frequency  
$\omega_c a /2 \pi c = 0.5$ (see Appendix B).  
The DOS exhibits a square-root singularity at the band edge  
$\omega_u a / 2 \pi c = 0.6$, as well as a UV divergence,  
$N_m(\omega) \propto \omega^2$, as $\omega \to \infty$; these are the   
characteristic features of this model. Due to the simultaneous presence of   
both divergences, this model clearly represents   
a severe numerical test of our approach. In order to test the method, 
we thus replace the PLDOS entering Eqs.\ (\ref{EqofMFin1}) and (\ref{EqofMFin2}) by  
$N_m(\omega)$.   
  
In Fig.\ 2, we present the results of our numerical solution for the   
radiation dynamics of a dipole oscillator with frequency $\omega_0$  
that is coupled to the modes of a PC, as described by   
Eqs.\ (\ref{EqofMFin1}) and (\ref{EqofMFin2}), for various values of  
the bare oscillator frequency, $\omega_0 a / 2 \pi c$, relative to the   
bandedge at $\omega_u a /2 \pi c=0.6$.  
The coupling strength has been chosen such that $g(\omega_0) = 10^{-4}$,  
corresponding to $\beta = 10^{-8} \times \omega_0^3$.   
  
Clearly visible are normal mode oscillations, also referred to as vacuum  
Rabi oscillations, and the fractional localization of the oscillator's energy  
at long times near the photonic band--edge \cite{Qua94}.  
As expected, for frequencies deep in the   
photonic band--gap ($\omega_0 a / 2 \pi c = 0.58$), where the system  
oscillator is effectively decoupled from the bath oscillators, we find no  
noticeable decay of the oscillator amplitude. Deep in the photonic conduction  
band ($\omega_0 a / 2 \pi c = 0.62$), the system oscillator is coupled to  
a bath with a smooth and slowly--varying mode density, as in free space.  We  
therefore observe exponential decay of the oscillator amplitude, though with  
a time scale that differs significantly from that in free space. Due to the  
large value of the DOS close to the photonic band edge, the initial decay is  
faster for bare oscillator frequencies close to this edge than for frequencies 
deep inside the allowed photonic band.  
These results were obtained for a smooth exponential cutoff for the DOS   
around $\Omega_c a/2 \pi c = 3.0$ and $M=2.5\times 10^5$ oscillators  
representing the modes of the PC. We also performed numerical simulations  
between all combinations of $\Omega_c$ and $M$ with values    
$\Omega_c a/2 \pi c = 3.0, 6.0, 9.0$ and $M= 2.5 \times 10^5,  
5 \times 10^5,10^6$ and found that the numerical values differ by at most  
0.2\% of the values shown in Fig 1. This demonstrates that, despite the  
presence of the singularities in the DOS, our approach still provides  
accurate and convergent results.  
   
\section{Discussion}   
In summary, we have developed a realistic field theory for an  
oscillating electric dipole located in a PC. The theory is based on the   
natural modes of the PC, the Bloch waves, and allows the direct   
incorporation of realistic band structure calculations in order to obtain   
quantitative results for the radiation dynamics of the dipole antenna.   
We have shown how the theory must be renormalized in order to account for   
unphysical divergences and have identified the classical analogue of the   
Lamb shift of the dipole's natural radiation frequency. Finally, we have    
developed a reliable numerical scheme based on a probability interpretation   
of the PLDOS that solves the equations of motion for the dipole oscillator  
coupled to the electromagnetic mode reservoir of the PC.  
  
The viability of this approach was demonstrated for an isotropic model DOS  
for which we have derived well-known results for radiating atomic systems  
\cite{Qua94} in the context of a radiating classical dipole.    
The model considered contains two divergences, one square-root-divergence at  
the photonic band edge and a quadratic UV-divergence, and therefore clearly   
comprises the most serious test of our approach.  
More realistic models of a three dimensional photonic band-edge take into  
account the anisotropy of the BZ, and therefore do not suffer from a  
band--edge singularity \cite{Vat98}.  As a result, our formalism is clearly  
more than capable of treating more realistic descriptions of the  
electromagnetic reservoir within a PC.  
  
Though we have developed our theory for an LC circuit 
in a microwave PC, we have pointed out in Section II that the formalism applies 
equally well to a 
semiclassical Lorentz  
oscillator model of an excited two-level atom.  Therefore, our 
approach is applicable to both microwave antenn\ae~ and to optical 
atomic transitions.  However, technological constraints suggest that 
microwave experiments will likely be easier to perform than optical 
experiments involving single atoms. 
As discussed, an appropriate microwave antenna could, for example,  
take the form of a high-Q metallic pin placed in or near a PC. The pin can  
then be excited by a focused ultrashort laser pulse that generates free  
carriers at one end;  
these carriers then undergo several oscillations across the pin before  
re--establishing charge equilibrium.  The resulting signal could be  
easily detected and compared with the emission from such an  
antenna positioned in free space, or within a homogeneous sample of  
the dielectric material that makes up the backbone of the PC  
under consideration.   
 
In its own right, such a microwave system could  
have considerable   
applications in radio science and microwave technology. For example, the   
PBG can be used as a frequency filter, and can be used to fine tune   
the bandwidth of a dipole emitter with a resonant frequency near the edge   
of the gap. It may also be possible to actively modify the photonic   
band structure, effectively changing the radiation pattern of a dipole   
emitter. A feasible scheme for active band structure modification has   
recently been proposed in the context of optical PCs \cite{Bus99}, in which   
the PC is   
infiltrated with a liquid crystalline material whose nematic director is   
aligned using applied electric fields. By rotating the director, it was   
found that the band structure could be significantly modified, and that   
PBGs may be opened and closed altogether. Similar methods   
may be applied to the case of microwave PCs.  
  
Although we have concentrated specifically on the linear model,  
the method of coupled oscillators can be extended to treat a nonlinear  
antenn\ae~, or a collection of two--level atoms in a regime where  
saturation effects arise.  
As we have shown here, this method of coupled classical oscillators  
reproduces effects normally associated with quantum optical calculations.  
We expect that a nonlinear oscillator model will reproduce some of  
the effects studied for a single two--level atom coupled to the  
modes of a PC without the need for quantizing the field.  
However, a classical treatment would need to be abandoned if  
multiphoton excitations are non--negligible \cite{Nik99}.  
Given that multiphoton effects are difficult to observe in  
the microwave domain\cite{Bru96} and even more challenging in the  
optical domain\cite{Car96}, it is reasonable to expect that a classical  
model of radiative dynamics in a PC should be sufficient for foreseeable  
experiments.  
  
\acknowledgments   
We are grateful to K.-J.\ Boller and R.\ Beigang for stimulating discussions   
concerning the experimental realization of radiating  
dipoles in microwave PCs.  
The work of KB the was supported by the Deutsche   
Forschungsgemeinschaft (DFG) under Grant Bu 1107/1-1.   
NV acknowledges support from the Ontario Graduate Scholarship Program.  
BCS acknowledges support from the Department of Physics,  
University of Toronto, and the support of an Australian Research  
Council Large Grant.   
This work was supported in part by the New Energy and Industrial Technology    
Development Organization of Japan and by Photonics Research Ontario.

\appendix   
   
\section{Classical field theory for Photonic Crystals}    
   
\label{sec:field}   
In this Appendix, we present a self-contained formulation of a classical
field theory for the Bloch modes of a PC, and we develop the Hamiltonian 
describing the coupling of a radiating dipole couples to these modes.  
As a first step, we   
review the computation of   
dispersion relations,    
and of electric and magnetic field modes from band structure calculations    
\cite{Bus98}. We then demonstrate how the results of such band structure    
calculations can be used to construct the corresponding vector    
potentials and free field Hamiltonian. Finally, we derive   
the full minimal coupling Hamiltonian for a classical radiating   
dipole embedded in a PC.   
This may be compared to the formulation of a general, quantized field
theory for EM modes in nonuniform dielectric media in terms of a normal
mode expansion in the context of quantum optics \cite{Gla91}.

\subsection{Review of band structure calculations}   
   
We develop our theory in terms of the magnetic   
field $\vec{H}$ rather than in terms of the electric or displacement    
fields because~(i)~$\nabla \cdot \vec{H} = 0$    
and,~(ii)~the transverse and longitudinal components of the    
magnetic field are continuous across the dielectric boundaries. This 
leads to more rapid convergence of the relevant Fourier series expansions.   
   
In a three-dimensional PC, we can write the   
eigenvalue equation for the magnetic field~$\vec{H}$ as   
\begin{equation}   
\label{DE:H}   
\nabla \times \left( \eta_p(\vec{r}) \, \nabla \times \vec{H} \right)   
	+ \frac{\omega^2}{c^2} \vec{H} = \vec{0}   
\end{equation}   
with~$\eta_p(\vec{r})$ the inverse of the periodic dielectric   
permittivity,   
\begin{equation}   
\label{epsilon_p}   
\epsilon_p(\vec{r}) = \epsilon_b + (\epsilon_a-\epsilon_b)   
\sum_{\vec{n} \in {\cal Z}^3} S(\vec{r}-\vec{n} \cdot \mbox{\boldmath $A$}).   
\end{equation}   
The medium is assumed to consist of a background material   
with bulk permittivity~$\epsilon_b$ and a set of scatterers,   
with bulk permittivity~$\epsilon_a$.  The shape of the scatterers is   
described by   
the function $S$, i.\ e.,, $S(\vec{r}) = 1$ if $\vec{r}$ lies inside   
the scatterer and zero elsewhere,    
distributed periodically at positions   
\begin{equation}   
\left\{ \vec{R} \right\} =    
\left\{ \sum_{i=1}^3 n_i \vec{a}_i | n_i \in {\cal Z} \right\}.   
\end{equation}   
The notation of Eq.\ (\ref{epsilon_p}) is obtained by defining the    
matrix~$\mbox{\boldmath $A$} =(\vec{a}_1 \, \vec{a}_2 \, \vec{a}_3)$   
and~${\cal Z}^3 = {\cal Z} \otimes {\cal Z} \otimes {\cal Z}$.   
The dielectric permittivity is spatially periodic modula   
$\vec{n} \cdot \mbox{\boldmath $A$}$.   
The assumption of a scalar permittivity is reasonable for   
bulk materials which are not birefringent but in no way restricts   
the considerations below. Chromatic dispersion effects are considered    
to be negligible, thus allowing the time-dependence of the permittivity to    
be ignored.   
Let us define the dual    
matrix~$\mbox{\boldmath $B$} = 2\pi (\mbox{\boldmath $A$}^{-1})^T$.   
For~$\mbox{\boldmath $B$}=(\vec{b}_1 \, \vec{b}_2 \, \vec{b}_3)$,   
this definition leads to the orthogonality relation   
\begin{equation}   
\vec{a}_i \cdot \vec{b}_j = 2\pi \, \delta_{ij} .   
\end{equation}   
Whereas the points $\vec{n} \cdot \mbox{\boldmath $A$} $ are the real space    
lattice vectors, the points $\vec{m} \cdot \mbox{\boldmath $B$}$, for    
$\vec{m} \in {\cal Z}^3$ are the reciprocal lattice vectors.   
The inverse permittivity can be expanded in the dual basis as   
\begin{equation}   
\eta_p(\vec{r}) = \sum_{\vec{m} \in {\cal Z}^3}   
\eta_{\vec{m}} e^{\imath \vec{m} \cdot \mbox{\boldmath $B$} \cdot \vec{r}} .   
\end{equation}   
   
The differential equation~(\ref{DE:H}) has periodic coefficients.   
By the Bloch-Floquet theorem we can expand the magnetic field as   
\begin{equation}   
\label{Floquet}   
\vec{H}_{\vec{k}} =    
e^{\imath \vec{k} \cdot \vec{r}} \, \vec{u}_{\vec{k}}(\vec{r})   
\end{equation}   
where~$\vec{u}_{\vec{k}}$ is spatially periodic modulo~$\mbox{\boldmath $A$}$;   
that is,    
\begin{equation}   
\vec{u}_{\vec{k}}(\vec{r})   
	= \vec{u}_{\vec{k}} (\vec{r} + \vec{n} \cdot \mbox{\boldmath $A$}) .   
\end{equation}   
The set $\{\vec{k}\}$ labeling the solutions can be restricted to lie    
within in the irreducible part of the first Brillouin zone (BZ), since    
any value of~$\vec{k}$ can then be obtained through a combination of   
group transformations   
with respect to an operation from the point group of the crystal   
and translations with respect to a reciprocal lattice vector.   
We can therefore express each wavevector $\vec{k}$ as   
\begin{equation}   
\vec{k} \equiv \vec{k}_{\mbox{\boldmath $T$},\vec{m}}    
        \equiv \vec{k}_* \cdot \mbox{\boldmath $T$}   
	+ \vec{m} \cdot \mbox{\boldmath $B$} ~~,   
\end{equation}   
where $\vec{k}_*$ is an element of the irreducible part of the 1. BZ   
and $\mbox{\boldmath $T$}$ an element of the crystal's point group.   
   
Applying the Bloch-Floquet theorem, Eq.\ (\ref{Floquet}), the magnetic    
field can be expanded as   
\begin{equation}   
\label{H:folded}   
\vec{H}_{\vec{k}} = e^{\imath \vec{k} \cdot \vec{r}}   
	\sum_{\vec{m}} \sum_{\lambda=1}^2 h_{\vec{m}}^{\vec{k},\lambda} \,   
	\hat{e}_{\vec{m}}^{\vec{k},\lambda} \,    
	e^{\imath \vec{m} \cdot \mbox{\boldmath $B$} \cdot \vec{r}} .   
\end{equation}   
Here $\lambda$ is the index of polarization and the vectors   
\begin{equation}   
\left\{ \hat{e}_{\vec{m}}^{\vec{k},1}, \hat{e}_{\vec{m}}^{\vec{k},2},   
	\frac{\vec{k}+\vec{m} \cdot \mbox{\boldmath $B$}}   
             {|\vec{k}+\vec{m} \cdot \mbox{\boldmath $B$} |} \right\}   
\end{equation}   
form an orthonormal right-handed triad.   
This expansion inserted into Eq~(\ref{DE:H}) yields an infinite eigenvalue   
problem which is then solved numerically by a suitable truncation.   
Typically the cardinality of the set~$\{ \vec{m} \}$ is on the order of    
$10^3$ \cite{Bus98}. For any given $\vec{k}_*$ we obtain a discrete set of    
eigenfrequencies $\omega_{n\vec{k}}$ and corresponding eigenfunctions    
$H_{n\vec{k}}$ which we label by the band index $n \in {\cal N}$.   
It is important to note that the expression for the electric field can be    
recovered from the magnetic field via   
\begin{equation}   
\label{EfromH}   
\vec{E}_{n \vec{k}} (\vec{r}) =    
- i \frac{c}{\omega_{n\vec{k}} \, \epsilon_p(\vec{r})}    
\nabla \times \vec{H}_{n\vec{k}} (\vec{r})   
\end{equation}   
In addition, the Bloch waves obey the following orthogonality relations:   
\begin{eqnarray}   
\label{Ortho:H}   
\int d^3r \,    
     \vec{H}_{n\vec{k}}^{*}(\vec{r}) \cdot   
     \vec{H}_{m\vec{k}^\prime}(\vec{r})   
                      & \propto &   
     \delta_{nm} \, \delta(\vec{k}-\vec{k}^\prime) ~~, \\   
\label{Ortho:E}   
\int d^3r \, \epsilon_p(\vec{r}) \,   
     \vec{E}_{n\vec{k}}^{*}(\vec{r}) \cdot   
     \vec{E}_{m\vec{k}^\prime}(\vec{r})   
                      & \propto &    
     \delta_{nm} \, \delta(\vec{k}-\vec{k}^\prime) ~~,   
\end{eqnarray}   
where the integration is over all space in both cases. We are free to   
choose the constants of proportionality in the above relations, and do   
so in the next subsection.   
   
   
\subsection{Free--field Hamiltonian}   
   
Based on the above considerations, we are now in a position to derive    
the general expressions for the scalar and vector potential, $\phi(\vec{r},t)$   
and $\vec{A}(\vec{r},t)$ respectively, for the classical Hamiltonian    
of the free radiation field. We find that the expressions become particularly   
transparent in the Dzyaloshinsky gauge, i.e., when $\phi(\vec{r},t) \equiv 0$.   
Then,   
\begin{eqnarray}   
\label{EfromA}   
\vec{E}(\vec{r},t) & = & -\frac{1}{c} \,    
  \frac{\partial \vec{A}(\vec{r},t)}{\partial \, t}  ,\\   
\label{HfromA}   
\vec{H}(\vec{r},t) & = & \nabla \times \vec{A}(\vec{r},t) ~~,    
\end{eqnarray}   
and the gauge condition    
$\nabla \cdot \left( \epsilon_p(\vec{r}) \vec{A}(\vec{r},t) \right) =0$,    
reveals that in a PC the natural modes of the    
radiation field are no longer transverse. This is of importance when    
quantizing the field theory \cite{Gla91,Law95}. Given Eqs.\ (\ref{DE:H}),    
(\ref{EfromH}), (\ref{EfromA}) and (\ref{HfromA}), it is now straightforward    
to derive the following expansion of the vector potential $\vec{A}(\vec{r},t)$    
\begin{eqnarray}   
\vec{A}(\vec{r},t) & = & \sum_n \int_{\rm BZ} \frac{d^3k}{(2\pi)^3}   
                   \sqrt{\frac{2\pi \xi c^2}{\omega_{n\vec{k}}}}*   
                   \nonumber \\   
\label{A}   
                   &   &  \, \, \, \, \, \, \, \, \, \, \left(   
                   \beta_{n\vec{k}}(t) \, \vec{A}_{n\vec{k}}(\vec{r})+   
                   \beta_{n\vec{k}}^*(t) \, \vec{A}_{n\vec{k}}^*(\vec{r})   
                   \right) ,   
\end{eqnarray}   
where the time evolution of the free field is described by   
$\beta_{n\vec{k}}(t) = \beta_{n\vec{k}}(0) e^{-\imath   
\omega_{n\vec{k}}t}$.    
The field modes $\vec{A}_{n\vec{k}}(\vec{r})$ obey   
\begin{equation}   
\nabla\times\nabla\times\vec{A}_{n\vec{k}}(\vec{r}) =   
\frac{\omega_{n\vec{k}}^2}{c^2} \, \epsilon_p(\vec{r}) \vec{A}_{n\vec{k}}(\vec{r}) ~~,   
\end{equation}   
which is the same equation as that for the electric field modes   
$\vec{E}_{n\vec{k}}(\vec{r})$ of Eq.\ (\ref{EfromH}). We now choose   
the normalization of $\vec{A}_{n\vec{k}}$ such that   
\begin{equation}   
   \int d^3r \, \epsilon_p(\vec{r}) \,    
   \vec{A}_{n\vec{k}}(\vec{r}) \cdot \vec{A}_{m\vec{k}^\prime}(\vec{r})   
   = \delta_{nm} \, \delta(\vec{k}-\vec{k}^\prime) ~~,    
\end{equation}   
\begin{eqnarray}   
   \int \, d^3r \,     
   \left(\nabla \times \vec{A}_{n\vec{k}}(\vec{r}) \right) \cdot    
   \left(\nabla \times \vec{A}_{m\vec{k}^\prime} (\vec{r}) \right)   
   & = & \nonumber \\   
   \frac{\omega_{n\vec{k}}^2}{c^2} \,   
   \delta_{nm} \, \delta(\vec{k}-\vec{k}^\prime) ~~.   
\end{eqnarray}   
This also fixes the normalization in Eqs.\ (\ref{Ortho:H}) and (\ref{Ortho:E}).   
As a consequence, the total electric and magnetic field are now given by   
\begin{eqnarray}   
\vec{E}(\vec{r},t) & = & \imath \sum_n \int_{\rm BZ} \frac{d^3k}{(2\pi)^3}   
                     \sqrt{\frac{2\pi \xi c^2}{\omega_{n\vec{k}}}}*   
                     \nonumber \\   
\label{exp:efield} &   &   
                     \, \, \, \, \, \, \, \, \, \, \, \,   
                     \left(   
                     \beta_{n\vec{k}}(t) \, \vec{E}_{n\vec{k}}(\vec{r})-   
                     \beta_{n\vec{k}}^*(t) \, \vec{E}_{n\vec{k}}^*(\vec{r})   
                     \right) ~,\\   
\vec{H}(\vec{r},t) & = & \sum_n \int_{\rm BZ} \frac{d^3k}{(2\pi)^3}   
                     \sqrt{\frac{2\pi \xi c^2}{\omega_{n\vec{k}}}}*   
                     \nonumber \\   
\label{exp:hfield} &   &   
                     \, \, \, \, \, \, \, \, \, \, \, \,     
                     \left(   
                     \beta_{n\vec{k}}(t) \, \vec{H}_{n\vec{k}}(\vec{r})+   
                     \beta_{n\vec{k}}^*(t) \, \vec{H}_{n\vec{k}}^*(\vec{r})   
                     \right) ~,   
\end{eqnarray}   
where  we have re--introduced the electric and magnetic field modes,    
$\vec{E}_{n\vec{k}}(\vec{r}) =    
    (\omega_{n\vec{k}}/c) \,\vec{A}_{n\vec{k}}(\vec{r})$ and   
$\vec{H}_{n\vec{k}}(\vec{r}) =    
    \nabla \times \vec{A}_{n\vec{k}}(\vec{r})$, respectively.   
Eqs.\ (\ref{exp:efield}) and (\ref{exp:hfield})   
finally lead us to the free field Hamiltonian    
\begin{equation}   
H_{\rm res} = \sum_n \int_{\rm BZ} d^3k \, \xi\omega_{n\vec{k}} \,     
            \vert \beta_{n\vec{k}} \vert^2 ~~.   
\end{equation}   
The only nonzero Poisson brackets are   
$ \left\{ \beta_{n\vec{k}}, \beta_{n\vec{k}}^* \right\} = \imath/w$.   
   
\subsection{Radiating dipole embedded in a Photonic Crystal}   
   
We consider the insertion of a point dipole    
into a PBG structure at a prescribed location $\vec{r}_0$. The   
free dipole oscillator is described by the Hamiltonian $H_{\rm dip}$   
\begin{equation}   
\label{freeosci}   
H_{\rm dip} = \frac{L \dot{q}^2}{2L} + \frac{1}{2} \, L\, \omega_0^2 \, q^2   
           = \xi\omega_0 \, \vert \alpha \vert^2 ,   
\end{equation}   
where the dipole's natural frequency is $\omega_0 =1/LC$ and the complex   
oscillator amplitude $\alpha$ is given in terms of the charge $q$ and   
``current'' $L \dot{q}$ as    
$\alpha(t) =    
   q(t) \, \sqrt{L\omega_0/2w} + \imath (L\dot{q}(t)) /\sqrt{2 \xi L \omega_0}$,   
with Poisson brackets $\left\{ \alpha, \alpha^* \right\} = \imath/\xi$.   
The point dipole couples to the electric field via its dipole   
moment $ d(t) = a q(t)$ with orientation $\hat{d}$,   
which yields the interaction energy    
\begin{equation}   
H_{\rm int} = - a q(t) \left(\hat{d} \cdot \vec{E}(\vec{r}_0,t) \right) .   
\end{equation}   
 
In the rotating wave approximation to the interaction   
term, the final minimal coupling Hamiltonian for a radiating    
dipole in a PC is   
\begin{equation}   
\label{Hamilton}   
H = H_{\rm dip}+H_{\rm res}+H_{\rm ct}+H_{\rm int} .   
\end{equation}   
Collecting all the above results we obtain   
\begin{eqnarray}   
H & = &  \xi\omega_0 \, \vert \alpha \vert^2 +   
         \sum_\mu \xi\omega_\mu \, \vert \beta_\mu \vert^2 + H_{\rm ct}    
         \nonumber \\   
  &   & - \imath \xi\sum_\mu    
          \left( \, \alpha^* \, g_\mu^* \, \beta_\mu -   
                    \alpha \, g_\mu   \, \beta_\mu^*    
          \right) .   
\end{eqnarray}   
Here, we have introduced the symbolic index $\mu \equiv (n\vec{k})$ and   
the coupling constants $g_\mu$   
\begin{equation}   
g_\mu \equiv g_\mu(\vec{r}_0) =    
ac \sqrt{\frac{\pi}{L \omega_0 \omega_\mu}}   
        \, \left(\hat{d} \cdot \vec{E}_\mu^*(\vec{r}_0) \right) .   
\end{equation}   
In addition, in Eq.\ (\ref{Hamilton}) we have introduced a mass    
renormalization counter term, $H_{\rm ct} = - \xi\Delta \, \vert \alpha 
\vert^2 $    
in order to cancel unphysical UV-divergent terms of our non-relativistic    
theory, as discussed in the main text.   
For completeness, we list here only the nonzero Poisson brackets and   
initial values for an initially excited radiating dipole coupled to the   
Bloch waves of a PC. This, together with the Hamilton    
function $H$ in Eq.\ (\ref{Hamilton}) completely defines our problem:   
\begin{equation}   
\left\{ \alpha, \alpha^* \right\} = \left\{ \beta_\mu, \beta_\mu^* \right\} =    
\frac{ \imath }{ \xi } ~~,   
\end{equation}    
where $\alpha(0) = 1$ and $\beta_\mu(0) = 0$ for all $\mu$.    
   
\section{Model dispersion relation and density of states}   
  
A particularly stringent test of our approach's ability to  
describe the dynamics of a radiating dipole in a PC  
comes from its application to a dipole coupled to a 3D isotropic photon  
dispersion model for the electromagnetic reservoir. In this model,  
the coherent scattering condition that characterizes the photonic band edge  
is assumed to occur at the same frequency  for all directions of  
propagation.  Clearly this is not the case in a real crystal, whose  
Brillouin zone cannot have full rotational symmetry.  As a result, the  
isotropic model overestimates the electromagnetic modes available at a  
band--edge, so that, for example, near the  
upper photonic band edge at   
frequency $\omega_u$, the corresponding DOS exhibits a divergence of  
the form    
$N(\omega) \propto 1/ \sqrt{\omega - \omega_u}$. Conversely,   
for large frequencies ($\omega \gg \omega_u$) the DOS will exhibit a   
UV-divergence, {\em i.e.}, $N(\omega) \propto \omega^2$, as is the case in  
free space. More realistic LDOS' coming from full  
photonic band structure computations do not suffer  
from the pathological band edge divergence of the isotropic model.  
However, by solving the model of a 3D isotropic photonic band   
gap, we make contact with previous results based on the isotropic  
model in the effective mass   
approximation \cite{Qua94}.  
  
Consider a 1D photonic dispersion relation in the extended zone  
scheme.  
In order to describe a PBG at wave number $k_0$ with  
central frequency $\omega_c = c k_0 = (\omega_u+\omega_l)/2$ and upper   
and lower band edge at $\omega_u$ and $\omega_l$, respectively, we   
use the following Ansatz  
\begin{equation}  
\label{modeldisp}  
\omega(k) = \left\{ \begin{array}{c}  
                   \omega_+ + c_+ \sqrt{(k-k_0)^2 + \gamma_+^2} ~~~  
                   \mbox{for} ~ k > 0 \\  
                   \omega_- + c_- \sqrt{(k-k_0)^2 + \gamma_-^2} ~~~  
                   \mbox{for} ~ k < 0   
                   \end{array} \right. ~~.  
\end{equation}  
Using the requirements   
$\omega(k=0)=0$,   
$\omega(k=k_0-0_+)=\omega_l$,  
$\omega(k=k_0+0_+)=\omega_u$,   
$\partial_k \omega(k=0) = \partial_k \omega(k\to \infty) = c$, and   
$\partial_k \omega(k=k_0-0_+) = \partial_k \omega(k=k_0+0_+) = 0$, the  
unknown parameters in (\ref{modeldisp}) can easily be expressed in terms of   
a single parameter $\eta = \omega_l/\omega_c$,  $1/2 < \eta \le 1$ that   
describes the size of the photonic bandgap, giving:  
$\omega_+ = \omega_c$, $c_+ = c$,   
$\gamma_+=k_0(1-\eta)$,   
$\omega_-= \omega_c (\eta^2)/(2\eta-1)$,   
$c_- = c \eta/\sqrt{2\eta-1}$, and   
$\gamma_- = k_0 (1-\eta)/\sqrt{2\eta-1}$.   
  
From the dispersion relation (\ref{modeldisp}), the corresponding DOS,  
{\em i.\ e.},  
$N(\omega) = \int d^3k \, \delta(\omega-\omega(k))$ is given by 
\begin{equation}  
\label{modeldos}  
N_m(\omega) = \left\{ \begin{array}{c}  
            4 \pi c_{-}^2  
\frac{ [ k_0 - \sqrt{(\omega - \omega_-)^2 / c_-^2 - \gamma_-^2} ]^2  
      (\omega_{-} - \omega)}  
     { \sqrt{(\omega - \omega_{-})^2 / c_{-}^2 - \gamma_{-}^2}} \\[2ex]  
\mbox{for} ~ 0 \le \omega \le \omega_l \\[2.5ex]   
            4\pi c_+^2   
\frac{ [ k_0+\sqrt{(\omega-\omega_+)^2/c_+^2 - \gamma_+^2} ]^2  
      (\omega-\omega_+)}  
     {\sqrt{(\omega-\omega_+)^2/c+-^2-\gamma_+^2}} \\[2ex]  
\mbox{for} ~ \omega_u \le \omega < \infty   
                   \end{array} \right.  
\end{equation}   
For sufficiently large gaps $(\eta \le 0.9)$ and bare eigenfrequencies  
$\omega_0$ of the radiating dipole close to the upper band edge, it is  
an excellent approximation to ignore the lower branch of the photon dispersion  
relation, {\em i.\ e.}, for $k \le k_0$. The resulting DOS for this so-called   
three-dimensional isotropic, one-sided bandgap model is shown in   
Fig. 1 for a value of gap width parameter $\eta=0.8$ and   
the gap center frequency $\omega_c a /2 \pi c = 0.5$. The square-root  
singularity   
at the band edge as well as the UV divergence $N_m(\omega) \propto \omega^2$  
as $\omega \to \infty$ are clearly visible.

\newpage
\clearpage
\noindent

Figure 1: 
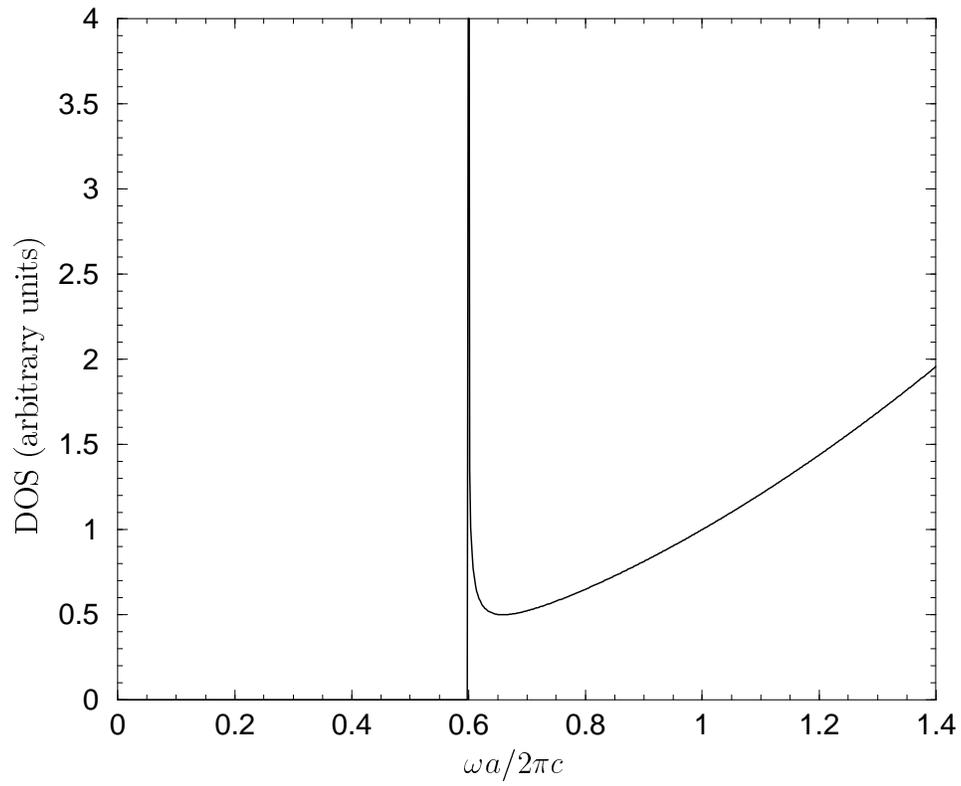
The DOS for the three-dimensional, isotropic one-sided bandgap 
        model of a PC. The parameters (see appendix B) are 
        $\eta=0.8$ and $\omega_c a /2 \pi c = 0.5$.\\
\ \\
\ \\
Figure 2:
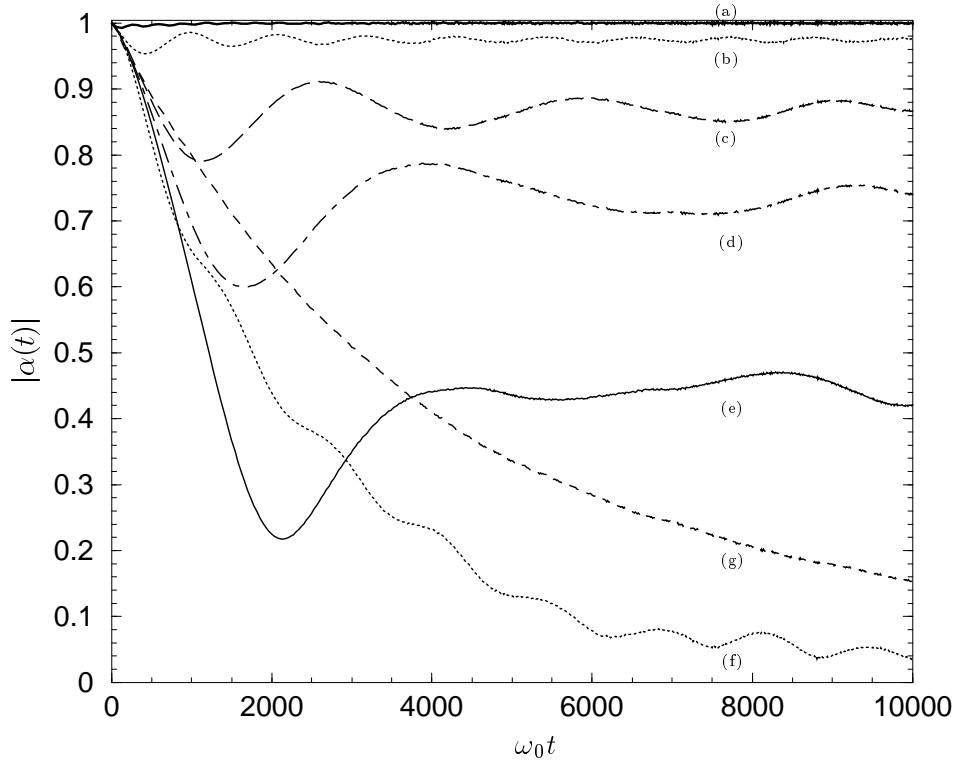
The radiation dynamics resulting from the three-dimensional, 
isotropic one-sided bandgap model DOS as shown in Fig. 1 for various 
values of the bare dipole oscillator frequency $\omega_0$ relative to 
the upper photonic bandedge $\omega_u$. The photonic bandedge is siutated 
at $\omega_u a / 2 \pi c = 0.6$ and the bare dipole oscillator frequencies
are  
(a) $\omega_0 a / 2 \pi c = 0.58$,  (b) $\omega_0 a / 2 \pi c = 0.595$,
(c) $\omega_0 a / 2 \pi c = 0.599$, (d) $\omega_0 a / 2 \pi c = 0.6$,
(e) $\omega_0 a / 2 \pi c = 0.601$, (f) $\omega_0 a / 2 \pi c = 0.605$, and  
(g) $\omega_0 a / 2 \pi c = 0.62$.
Clearly visible are normal--mode oscillations, or vacuum Rabi oscillations, 
and the fractional localization of radiation near the photonic bandedge. 
The coupling strength has been chosen such that $g(\omega_0) = 10^{-4}$. 
For frequencies deep in the photonic bandgap 
($\omega_0 a / 2 \pi c = 0.58)$ and deep in the photonic conduction band 
($\omega_0 a / 2 \pi c = 0.62)$ we observe negligible and exponential 
decay, respectively.
   
\end{document}